%% file: acl26.tex
\documentclass[11pt]{article}

\usepackage[final]{acl}

\usepackage{times}
\usepackage{latexsym}

\usepackage[T1]{fontenc}
\usepackage[utf8]{inputenc}

\usepackage{microtype}

\usepackage{inconsolata}

\usepackage{graphicx}

\usepackage{amsmath,amsfonts}
\usepackage{graphicx}
\usepackage{textcomp}
\usepackage{tabularx}
\usepackage{xcolor,soul}
\def\BibTeX{{\rm B\kern-.05em{\sc i\kern-.025em b}\kern-.08em
    T\kern-.1667em\lower.7ex\hbox{E}\kern-.125emX}}

\usepackage{booktabs}   %% For formal tables:
\usepackage{subcaption} %% For complex figures with subfigures/subcaptions
\usepackage{array}
\usepackage{ragged2e}
\usepackage{float}
\usepackage{listings}
\usepackage{xspace}
\usepackage{multirow}
\usepackage{amsthm}
\usepackage{balance}

\usepackage[skins]{tcolorbox}

\usepackage{xcolor,pifont}
\newcommand*\colourcheck[1]{%
	\expandafter\newcommand\csname #1check\endcsname{\textcolor{#1}{\ding{52}}}%
}

\usepackage{enumitem}
\usepackage[normalem]{ulem} % Provides dotted underline with \dotuline, disables \emph change
\usepackage{array}
\usepackage{wrapfig}
\usepackage{amsmath,amsfonts}
\usepackage[noend]{algpseudocode}
\usepackage{graphicx}
\usepackage{textcomp}
\usepackage{float}
\usepackage{listings}
\usepackage{xspace}
\usepackage{multirow}
\usepackage{amsthm}

\usepackage{balance}
\usepackage{algorithm}
\usepackage{algpseudocode}
\usepackage{colortbl}
\usepackage{subcaption}
\usepackage[skins]{tcolorbox}
\usepackage{xcolor}
\usepackage{multicol}
\usepackage{soul}
\usepackage{framed}
\usepackage{booktabs}
\usepackage{booktabs}
\usepackage{multirow}
\usepackage{siunitx}
\usepackage{minted}
\usepackage{multicol}
\usepackage{fancyvrb}
\usepackage{tcolorbox}
\colourcheck{blue}
\colourcheck{green}
\colourcheck{red}
\definecolor{custom-blue}{rgb}{0,0,1}
\definecolor{custom-red}{rgb}{1,0,0}

\newtcolorbox{boxB}[2][]{%
  enhanced,colback=white,colframe=black,coltitle=black,
  sharp corners,
  toprule=1.0pt,
  rightrule=0.3pt,
  leftrule=0pt,
  bottomrule=0pt,
  fonttitle=\itshape\scshape\large,
  left=0pt,right=5pt,top=5pt,bottom=3pt,
  attach boxed title to top right={yshift=-0.3\baselineskip-0.4pt,xshift=-5mm},
  boxed title style={tile,size=minimal,left=0.2mm,right=0.5mm,
    colback=white,before upper=\strut},
  title=#2,#1
}
\definecolor{darkred}{rgb}{0.6,0,0}

\newcommand{\tool}{\textsc{W-RAG}\xspace}
\title{W-RAG: Source-Aware Retrieval for Enterprise Document Generation from Heterogeneous Knowledge Bases}

\author{
  Hridya Dhulipala$^{1}$ \quad
  Rajesh Ombase$^{2}$ \quad
  Michael Wang$^{3}$ \quad
  Tien N. Nyugen$^{1}$ \\
  $^{1}$The University of Texas at Dallas \\
  $^{2}$DIGITAL MANAGEMENT, LLC (DMI) \\
  $^{3}$Massachusetts Institute of Technology \\
  \texttt{\{hridya.dhulipala, tien.n.nyugen\}@utdallas.edu} \\
  \texttt{rombase@dmigs.com} \\
  \texttt{wangrzm@mit.edu}
}

\begin{document}
\maketitle
\begin{abstract}
Retrieval-Augmented Generation (RAG) enables large language models to incorporate external knowledge during generation, improving factual grounding and domain adaptability. However, existing RAG pipelines assume that evidence retrieved from multiple repositories can be ranked globally using a single similarity function. While suitable for open-domain retrieval, this assumption breaks down in enterprise document generation, where heterogeneous knowledge bases (such as policies, regulations, technical documentation, and departmental guidelines) serve distinct roles and must be jointly represented in the generated document. As a result, global ranking often produces unbalanced context dominated by a subset of sources, leading to incomplete enterprise drafts. To address this limitation, we propose W-RAG, a source-aware retrieval framework that performs ontology-guided retrieval, local ranking within each knowledge base, and source-level weighting to regulate evidence composition. We further introduce a new dataset for retrieval-grounded enterprise document generation spanning multiple document types and industry domains. Experiments show that standard RAG pipelines struggle on this task, while W-RAG significantly improves document coverage and generation quality.
\end{abstract}

\input{sections/intro-3}

\input{sections/related-works}

\input{sections/methodology}

\input{sections/dataset}

\input{sections/experiment-setup}

\input{sections/conclusion}

\input{sections/limitations}

% \input{sections/evaluation-metrics}

% Bibliography entries for the entire Anthology, followed by custom entries
%\bibliography{custom,anthology-overleaf-1,anthology-overleaf-2}

% Custom bibliography entries only
\bibliography{sections/references}

% \newpage
\appendix
\section{Appendix}
\input{sections/appendix}

\end{document}

%% file: sections/intro-3.tex
\section{Introduction}
\label{sec:intro-3}

Retrieval-Augmented Generation (RAG) has emerged as a powerful paradigm for grounding large language models in knowledge, significantly improving factuality, transparency, and domain adaptability~\cite{lewis2021retrievalaugmentedgenerationknowledgeintensivenlp, gao2024retrievalaugmentedgenerationlargelanguage, guu2020realmretrievalaugmentedlanguagemodel}. By combining neural generation with targeted retrieval from repositories, RAG systems help organizations synthesize policies, contracts, technical documentation, and client-facing materials with greater reliability than purely parametric models~\cite{jadadgarcia2024foundationscomputationalmanagementsystematic}.

Despite this progress, current enterprise RAG implementations still inherit structural limitations. Recent studies and experience reports suggest that naïve, similarity-only RAG pipelines often underperform on document-generation tasks such as policy creation, proposals, Request for Information documents (RFIs), and compliance reports in heterogeneous corporate settings ~\cite{bruckhaus2024ragdoesworkenterprises, hadfield2026regulatorymarketsfutureai, engineeringgap, infosysenterpriseragai, brynjolfsson2024generativeaiwork, jadadgarcia2024foundationscomputationalmanagementsystematic}. This is because existing RAG systems retrieve candidates from multiple heterogeneous knowledge bases and then rank them globally before selecting the top-$k$ context for generation. While this design is natural in open-domain retrieval, it is problematic for enterprise document generation because it assumes that evidence from heterogeneous repositories can be compared using a single ranking function. In practice, however, enterprise knowledge bases differ in scope, authority, density, and role. Previous policies, regulations, scientific knowledge, playbooks, department notes, and technical specifications should be considered differently
%not all compete in one shared ranking pool without 
with regard to their influence in the drafting process. Global ranking tends to favor knowledge bases that are larger, denser, or more lexically aligned with the query, while suppressing smaller but necessary repositories. The result is often a context window dominated by one source family, with critical evidence from other departments or domains omitted. For enterprise document generation, the problem is thus not only ranking quality at the chunk level, but evidence composition at the document level.

In this paper, we propose {\tool}, a retrieval framework for enterprise document generation that improves both evidence selection and source composition across multiple knowledge bases. {\tool} operates in three stages. First, it uses ontology-guided topic modeling to identify passages that are not only lexically similar to the query, but also more aligned with the thematic requirements of the target document. Second, it ranks evidence locally within each knowledge base instead of collapsing all retrieved candidates into a single global ranking. Third, it uses source-level (or user-defined) \emph{weights} to regulate how much context is selected from each knowledge base, so that the final evidence set better reflects document needs rather than the dominance of any one repository.

Additionally, a major obstacle in studying this problem is the lack of a dataset for retrieval-grounded enterprise document generation. Existing RAG datasets primarily evaluate question answering ~\cite{yang2018hotpotqadatasetdiverseexplainable, Chen_Lin_Han_Sun_2024, karpukhin-etal-2020-dense, kwiatkowski-etal-2019-natural, joshi2017triviaqalargescaledistantly, rajpurkar-etal-2016-squad, bajaj2018msmarcohumangenerated, trischler-etal-2017-newsqa}, summarization, or general knowledge-intensive generation~\cite{berant-etal-2013-semantic}, and do not capture the multi-source evidence composition required in enterprise writing. To address this gap, we introduce a {\bf new dataset} covering multiple enterprise document types, including policies, product launch documents, M\&A due diligence reports, and academic program proposals. The benchmark spans several industry domains, including AI Technology \& Digital Media, Healthcare \& Telehealth, Climate \& Sustainabilty and Finance. It is designed so that successful generation requires retrieving and combining evidence from heterogeneous knowledge bases rather than relying on a single semantically aligned source.

Using this dataset, we evaluate multiple open-source and proprietary LLM-based systems under standard RAG configurations and find that they perform poorly on enterprise document generation despite producing fluent outputs. Our analysis shows that the generated enterprise document is sensitive to the knowledge bases represented in retrieved context. When retrieval is performed through global ranking alone, it often leads to drafts that fail to cover the full set of requirements specified by the requested enterprise document.

Our contributions are threefold:

(1) \textbf{Problem formulation}. We identify enterprise document generation as a retrieval setting in which standard RAG assumptions are poorly matched to task requirements, and characterize a key failure mode of existing systems: global ranking across heterogeneous knowledge bases, which fails to preserve the source composition needed for complete document generation.

(2) \textbf{Method.} We propose a source-aware retrieval framework for enterprise document generation, based on weighted local retrieval and supported by expert-authored ontological guidance, to improve cross-domain evidence selection and context composition.

(3) \textbf{Dataset and findings}. We introduce a new dataset for retrieval-grounded enterprise document generation across multiple document types and domains, and show that existing RAG algorithms underperform on this dataset, while our source-aware retrieval approach yields substantial improvements.

%% file: sections/related-works.tex
\section{Related Works}
\label{sec:related-works}

\paragraph{Retrieval-Augmented Generation}

Retrieval-Augmented Generation (RAG) has become a cornerstone of knowledge-grounded language modeling by integrating external retrieval with generative reasoning. The original RAG framework by Lewis et al.~\cite{lewis2021retrievalaugmentedgenerationknowledgeintensivenlp} and subsequent models such as REALM~\cite{guu2020realmretrievalaugmentedlanguagemodel, izacard-grave-2021-leveraging}, and RETRO~\cite{borgeaud2022improvinglanguagemodelsretrieving} showed that retrieving textual evidence before generation improves factuality and reduces hallucination. More recent works~\cite{gao2024retrievalaugmentedgenerationlargelanguage} extend these paradigms by improving retriever–generator interaction through adaptive evidence routing~\cite{DBLP:conf/coling/ZhengLLLLW24, li2026evirerankadaptiveevidenceconstruction}, and reinforcement learning from human  feedback~\cite{zhang2025opengenalignpreferencedatasetbenchmark, li2025r3raglearningstepbystepreasoning}.
Despite these advances, most RAG pipelines still assume homogeneous and open-domain corpora, relying on dense or hybrid similarity retrieval to select support documents. Such an assumption becomes problematic in enterprise or multi-source environments, where repositories differ substantially in authority, structure, and linguistic domain~\cite{bruckhaus2024ragdoesworkenterprises, arrieta2019explainableartificialintelligencexai}.

\paragraph{Multi-Source and Structured Retrieval}
A growing body of work has explored retrieval beyond single-source settings. Multi-hop and multi-granularity retrieval methods~\cite{yang2018hotpotqadatasetdiverseexplainable, tang2024multihopragbenchmarkingretrievalaugmentedgeneration, linjiaen-etal-2025-optimizing} aim to combine evidence across documents, yet they generally index homogeneous, task-specific datasets such as Wikipedia or news archives. Several studies on domain-specific or vertical RAG systems have sought to address this heterogeneity. Works in legal~\cite{kabir2025legalraghybridragmultilingual, reuter2025reliableretrievalragsystems,
butler2026legalragbenchendtoend}, biomedical~\cite{10822837,
11176078,
Matsumoto2024KRAGENAK}, and financial domains~\cite{george-etal-2025-enhancing,
10.1145/3746252.3761643,
wang2025financialanalysisintelligentfinancial} introduce hierarchical or ontology-based retrieval pipelines to better reflect domain semantics. However, these approaches focus on accuracy within one tightly scoped corpus. Enterprise applications, however, often require synthesizing information dispersed across knowledge bases that play complementary roles. Empirical evaluations from industrial deployments~\cite{bruckhaus2024ragdoesworkenterprises, infosysenterpriseragai} confirm that global top-k retrieval leads to skewed evidence selection, where large or lexically dominant sources overshadow smaller yet essential ones.

\paragraph{Datasets for Knowledge-Grounded Generation}
Benchmark datasets for RAG typically target question answering~\cite{karpukhin-etal-2020-dense, kwiatkowski-etal-2019-natural, joshi2017triviaqalargescaledistantly, rajpurkar-etal-2016-squad, bajaj2018msmarcohumangenerated}, fact verification~\cite{sorodoc2025garagebenchmarkgroundingannotations}, or short-form summarization~\cite{berant-etal-2013-semantic}. Multi-document reasoning datasets like HotpotQA~\cite{yang2018hotpotqadatasetdiverseexplainable} encourage evidence chaining but still draw from single knowledge domains. Recent datasets for long-form factual generation~\cite{zhao2024longragdualperspectiveretrievalaugmentedgeneration}, safety policy writing~\cite{hadfield2026regulatorymarketsfutureai}, and organizational compliance~\cite{jadadgarcia2024foundationscomputationalmanagementsystematic} indicate emerging interest but lack heterogeneous knowledge-source coverage. Our introduced dataset extends this direction by emphasizing cross-domain evidence composition, serving as a benchmark for evaluating retrieval-weighted RAG methods in enterprise contexts.

%% file: sections/methodology.tex
\section{Method}
\label{sec:methodology}

\subsection{Overview}

\textsc{{\tool}} is a source-aware retrieval framework for enterprise document generation over heterogeneous knowledge bases (KBs). We use the term \emph{context allocation} to refer to the token-level distribution of the final prompt context across KBs. Given an input query or seed document $q$, a total retrieval budget $T_{\text{total}}$, and $M$ KBs, the goal is to construct a context window by selecting text chunks from each KB such that the total number of retrieved tokens does not exceed $T_{\text{total}}$ while the retrieved evidence remains relevant to the target drafting task.

At a high level, \textsc{{\tool}} proceeds in three steps: (1) it extracts ontology-grounded topics from the input and retrieves candidate chunks from each KB, (2) it estimates and recommends how much of the final context should come from each KB, and (3) it converts these source preferences into token budgets and builds a single source-balanced context window for generation. 

\subsection{Stage 1: Ontology-Guided Candidate Retrieval}

\textsc{{\tool}} first extracts a structured topic representation of the input using an LLM guided by a domain ontology. An ontology is a formal representation of the key concepts in a domain and the relations among them. Unlike a taxonomy, an ontology allows richer, non-hierarchical relations between entities, concepts, and categories. In our system, the ontology is used both to organize domain concepts and to map free-form input into a structured topic set that can guide retrieval. For example, in a policy drafting task, extracted topics such as \emph{data privacy}, \emph{access control}, and \emph{compliance reporting} may be mapped to ontology categories such as \emph{security} and \emph{governance}. Details of ontology construction and topic-extraction prompts are provided in~\ref{sec:appendix-ontology}.

Given input $q$, the topic extractor produces
\[
T=\{(t_i,g_i)\}_{i=1}^{N},
\]
where $t_i$ is an extracted topic phrase and $g_i$ is its ontology category.

Each KB $j$ is indexed as a collection of text chunks (e.g., paragraph-level segments). We denote the $k$-th chunk from KB $j$ by $c_{jk}$. At indexing time, each chunk is also associated with a set of extracted ontology topics, denoted by $M(c_{jk})$. \textsc{{\tool}} retrieves candidate chunks from each KB and scores them using a hybrid relevance function:
\begin{equation}
R(c_{jk}) = f_{\text{sim}}(q, c_{jk}) + \lambda\, f_{\text{ont}}(T, c_{jk}),
\label{eq:hybrid-score}
\end{equation}
where $f_{\text{sim}}$ measures semantic similarity between the input and the chunk, $f_{\text{ont}}$ measures ontology-based topic alignment, and $\lambda \ge 0$ controls the relative influence of ontology guidance.

Semantic similarity is computed as cosine similarity between dense embeddings:
\begin{equation}
f_{\text{sim}}(q, c_{jk}) = \cos\big(\phi(q), \phi(c_{jk})\big),
\end{equation}
where $\phi(\cdot)$ is the embedding function used by the retriever.

To compute ontology alignment, we compare each input topic against the ontology topics associated with the chunk. For an input topic $t_i$ and indexed chunk topic $m \in M(c_{jk})$, we use the rule-based match score
\[
s(t_i,m)=
\begin{cases}
1.0 & \text{if $t_i$ exactly matches $m$,}\\
0.7 & \text{if one string contains the other,}\\
0.3 & \text{only ontology category match,}\\
0 & \text{otherwise.}
\end{cases}
\]
These values were selected to enforce a clear ordering of match strength (exact \> partial \> categorical \> none) while keeping the scoring scheme simple and interpretable. 
Thus, chunk-level ontology score is : 
\begin{equation}
f_{\text{ont}}(T,c_{jk}) =
\frac{1}{N}\sum_{i=1}^{N}\max_{m \in M(c_{jk})} s(t_i,m).
\label{eq:font}
\end{equation}
Thus, $f_{\text{ont}}$ is the average best-match score between the input topics and the chunk's indexed ontology topics. In practice, chunks that mention more of the input topics---or mention them more specifically---receive higher ontology scores.

In all experiments, we set $\lambda=0.5$. We selected this value on a held-out development set by sweeping $\lambda \in [0,1]$ and choosing the value that improved average requirement satisfaction without degrading context relevancy. The same value is used across all domains and document types.

For each KB $j$, we sort chunks by $R(c_{jk})$ and retain the top-$m$ candidates:
\begin{equation}
C_j = \{c_{j1}, c_{j2}, \ldots, c_{jm}\},
\end{equation}
where $C_j$ denotes the top-$m$ candidate text chunks retrieved from KB $j$ under the hybrid score in Eq.~\ref{eq:hybrid-score}. These per-KB candidate sets are then used for source weighting and final context construction.

\subsection{Stage 2: Source Weight Estimation}

Stage~1 identifies thematically relevant candidates, but it does not yet determine how much each KB should contribute to the final prompt. A standard global top-$k$ strategy would simply merge all candidates into one shared ranking, which can over-concentrate the final context on a single source. \textsc{{\tool}} instead estimates a source weight for each KB before constructing the final context.

For each KB $j$, we maintain a KB-level topic index $M_j$, obtained by aggregating the ontology topics associated with documents in that KB. Source weighting is computed in real time by comparing the input topic set $T$ against these KB topic indexes. The main source signal is topical overlap:
\begin{equation}
S^j_{\text{topic}} =
\frac{1}{N}\sum_{i=1}^{N}\max_{m \in M_j} s(t_i,m),
\label{eq:stopic}
\end{equation}
where the same exact / partial / category-only match function $s(\cdot,\cdot)$ is used as in Stage~1. Thus, $S^j_{\text{topic}} \in [0,1]$ measures how well KB $j$ matches the input topics overall.

We combine $S^j_{\text{topic}}$ with two auxiliary terms. The first is a weak document-availability prior:
\[
S^j_{\text{doc}} = \min\left(1, \frac{N^j_{\text{doc}}}{100}\times 0.5\right),
\]
where $N^j_{\text{doc}}$ is the number of indexed documents in KB $j$. This term gives a small boost to KBs with broader supporting coverage while preventing size alone from dominating the score. The second is a category bonus:
\[
S^j_{\text{cat}} =
\begin{cases}
0.2 & \text{KB $j$ matches at least 1 input topic,}\\
0 & \text{otherwise.}
\end{cases}
\]
This term rewards coarse domain alignment even when exact topic overlap is limited.

The final source score for KB $j$ is
\begin{equation}
S^j_{\text{final}} = 0.7S^j_{\text{topic}} + 0.1S^j_{\text{doc}} + 0.2S^j_{\text{cat}}.
\label{eq:sfinal}
\end{equation}
The coefficients reflect the intended priority of the three signals: topical overlap is the primary driver, category alignment provides domain-level support, and document availability acts only as a weak prior. We fixed these coefficients after preliminary development-stage tuning and kept them constant across all experiments. The resulting scores are normalized into percentage-based KB proportions (termed as \emph{weights}):
\begin{equation}
w_j = \frac{S^j_{\text{final}}}{\sum_{k=1}^{M} S^k_{\text{final}}}\times 100,
\qquad
\sum_{j=1}^{M} w_j = 100.
\label{eq:weights}
\end{equation}

This procedure is computed in real time: the recommender does not rely on static precomputed KB preferences, but instead compares the current input topics against the current topic indexes of the KBs. When new documents are added to a KB, they affect future recommendations only after re-indexing updates the associated topic index and document counts.

\subsection{Stage 3: Weighted Context Construction}

Since the generator has a fixed context window, the KB weights in Eq.~\ref{eq:weights} must be converted into token budgets. For each KB $j$, we define
\begin{equation}
T_j = \frac{w_j}{100}T_{\text{total}},
\qquad
\sum_{j=1}^{M} T_j = T_{\text{total}}.
\label{eq:token-budget}
\end{equation}

Weights are defined as \emph{the proportional allocation of context tokens across knowledge bases based on their estimated relevance to the query.}

These budgets specify how many tokens each KB is allowed to contribute to the final prompt context. 

\textsc{{\tool}} then selects the highest-ranked chunks from each locally ranked candidate set $C_j$ until the corresponding budget $T_j$ is exhausted. The selected chunks are merged into a Single Context Window (SCW), and each chunk is annotated with its source KB to preserve provenance. This step is the mechanism that enforces source-level constraints on the final context window, which standard RAG pipelines do not control. In this way, \textsc{{\tool}} transforms source relevance estimates into a generation context whose composition reflects the informational needs of the target document rather than the dominance of a single source.

The final prompt is formed as
\[
P = \{I_{\text{sys}}, q, \text{SCW}\},
\]
where $I_{\text{sys}}$ denotes system instructions. This prompt is then passed to the LLM for enterprise document generation.

% \paragraph{Complexity and implementation remarks.}
% The additional cost of \textsc{{\tool}} beyond standard RAG comes from per-KB topic matching and per-KB local ranking. In our setting, this cost scales linearly with the number of KBs and the number of indexed topics examined per KB. Because the number of KBs is small ($M=4$ in our benchmark) and KB topic indexes are queried online with a capped retrieval size, this overhead is modest relative to generation latency. Prompts for topic extraction and other implementation details are provided in Appendix~\ref{app:ontology-prompts}.

%% file: sections/dataset.tex
\section{Dataset}
\label{sec:dataset}

Evaluating Retrieval-Augmented Generation (RAG) for enterprise-level document generation presents a unique challenge, as no standardized dataset currently exists for long-form, requirement-driven document synthesis. To address this gap, we introduce a novel dataset designed to evaluate RAG techniques for enterprise document generation under heterogeneous knowledge base conditions. Each data sample consists of an input requirement document that specifies detailed constraints and objectives for generating a target document. A sample of an input document is shown in the~\ref{sec:appendix-dataset} 

The dataset contains 100 requirement input documents. Each requirement document was manually constructed and belongs to one of four enterprise document types: \emph{Policies, Product launch documents, Academic program documents \& Merger and acquisitions due diligence documents}. These document types were selected because they represent distinct enterprise requirements with different evidence needs, rhetorical structures, and decision-making purposes. Details on the how the input documents were created and authored can be found in~\ref{sec:appendix-dataset}.

Across these document types, the dataset covers four high-level topical themes: \emph{AI Technology \& Digital Media, Healthcare \& Telehealth, Climate \& Sustainabilty and Finance}. These themes were chosen to ensure that the benchmark reflects realistic enterprise themes in which document generation must integrate business, technical, legal, and market-specific information.

To support retrieval, we construct four heterogeneous knowledge bases (KBs), each containing approximately 500 publicly sourced documents collected from open-access web resources:

(1) \textbf{KB1 (Corporate)}: Corporate filings, company reports, operational documentation, and governance materials.~\cite{edgar_sec,alpha_vantage}

(2) \textbf{KB2 (Scientific)}: Peer-reviewed journal articles, technical whitepapers, and research summaries. ~\cite{arxiv, ncbi_nih}

(3) \textbf{KB3 (Regulatory)}: Legal texts, compliance guidelines, and regulatory frameworks. \cite{fda_gov, sec_gov}

(4) \textbf{KB4 (Market \& News)}: Market surveys, industry reports, and news articles. ~\cite{forbes}

The goal of the using such knowledge bases is to test RAG systems' ability to assemble a context that reflects the different functional roles of enterprise evidence. For instance, a product launch document in healthcare AI may need corporate precedent, scientific validation, regulatory requirements, and market positioning simultaneously.

%% file: sections/experiment-setup.tex
\section{Experiments}
\label{sec:experiment-setup}

\subsection{Baselines \& Metrics}

\paragraph{Baselines}
We compare {\tool} against the following representative RAG pipelines: %to isolate the impact of????

(1) \emph{Vanilla RAG.} The standard RAG pipeline~\cite{lewis2021retrievalaugmentedgenerationknowledgeintensivenlp} that performs dense semantic retrieval over all available knowledge bases (KBs), followed by a global ranking of retrieved documents across KBs. The top-ranked passages are then provided to the generator without any explicit modeling of domain structure or source-level preferences.

(2) \emph{OG-RAG}. The second baseline augments retrieval with ontology-guided topic modeling~\cite{sharma-etal-2025-og}.  For each of the four domains, an ontology was manually curated capturing the major concepts, entities, and thematic relations relevant to document drafting in that domain. Retrieved documents are first filtered and organized using this ontology-guided topic representation, after which passages are globally ranked across KBs and passed to the generator. OG-RAG was shown to demonstrate higher recall of accurate facts and response correctness compared to matrix (like RAPTOR~\cite{sarthi2024raptorrecursiveabstractiveprocessing}) and graph (like GraphRAG~\cite{edge2025localglobalgraphrag}) based RAG baselines, establishing ontology-guided retrieval as a strong foundation.

\paragraph{Ontologies}
\label{sec:create-ontology}

We construct one ontology for each domain--document-type pair, resulting in 16 ontologies overall. Each ontology is built using a semi-automated process, where an initial automatically generated version is manually refined and verified by domain experts. These ontologies support topic extraction and ontology-guided retrieval in OG-RAG and {\tool}.

% Our proposed method, , extends OG-RAG by incorporating prescribed KB weights during retrieval and evidence aggregation. In addition to ontology-guided topic modeling, {\tool} explicitly controls the relative contribution of different KBs, allowing the system to better align retrieval and generation with the intended evidence distribution.

% \subsection{Evaluation Metrics}
% Because document generation quality is multi-dimensional and largely subjective in nature, we employ three complementary evaluation metrics:

% Answer Relevancy (LLM-as-Judge).
% We use a separate evaluation model operating in judge mode to assess how well the generated document satisfies the input requirements. The judge model scores outputs along requirement coverage, coherence, and constraint adherence. This provides a holistic quality estimate beyond lexical overlap.

% Context Relevancy.
% We compute cosine similarity between the embedding of the input requirement document and the aggregated embeddings of retrieved contexts. This measures how semantically aligned the retrieved evidence is with the task specification.

% Knowledge Base Influence (KBI).
% We introduce Knowledge Base Influence, a metric quantifying the extent to which generated content is grounded in retrieved knowledge rather than parametric hallucination. Using sentence-level attribution, each generated sentence is aligned with its most similar retrieved chunk. KBI measures:

% The proportion of sentences supported by retrieved context.

% The average semantic similarity between generated sentences and their attributed evidence.

\paragraph{Evaluation Metrics}
Due to the multi-dimensional nature of enterprise document generation, we employ the following complementary evaluation metrics:

(1) \textbf{\emph{Requirement Satisfaction}} is defined as the extent to which a generated document satisfies the minimum essential requirements outlined in the input document. Unlike QA benchmarks, where correctness can often be judged against a single reference answer, enterprise documents are better evaluated against a requirement checklist specifying the mandatory elements that must appear in a valid draft.

For each document type (e.g., policy documents, product launch briefs, M\&A due diligence summaries), we construct a checklist of essential requirements in consultation with domain practitioners who routinely author such documents and each checklist contains approximately 12 core requirements. Each requirement is scored on a 2-point ordinal scale: \textbf{0: Not satisfied, 1: Partially satisfied / insufficiently specified \& 2: Fully satisfied}. This results in a maximum score of 24 points per document. The score is then normalized. We semi-automate the scoring process by using an \textit{LLM-as-a-judge} that is provided with:

(1) The input document or drafting prompt, 

(2) The generated output and 

(3) The requirement checklist for the corresponding document type.

The judge is instructed to assign a score for each checklist item. To improve reliability, these scores are then manually verified by domain practitioners before being used in analysis. The standard instructions given to domain practioners to validate the scoring is provided in~\ref{sec:appendix-manual-validation}.

(2) \textbf{\emph{Context Relevancy}} measures the proportion of retrieved passages that are genuinely relevant to the input drafting task.

Given a set of $N$ retrieved passages $\{p_i\}_{i=1}^{N}$, an LLM judge labels each passage as relevant or non-relevant with respect to the input document and drafting objective. Let $z_i \in \{0,1\}$ denote this judgment. The Context Relevancy is defined as:

\[
\text{CTS} =
\frac{1}{N}
\sum_{i=1}^{N} z_i
\]

As with Answer Relevancy, LLM-based judgments are manually validated to ensure scoring consistency.

Context Retrieval as a metric isolates retrieval quality from generation quality: a system may generate fluent text, but poor retrieval will often limit factual adequacy and source grounding.

(3) \textbf{\emph{Realized Knowledge Base Fidelity}} The goal of this metric is to estimate which knowledge base most likely supports different parts of the generated document, rather than to detect verbatim reuse of retrieved passages. Because modern LLMs typically synthesize and paraphrase retrieved evidence, attribution is performed using semantic similarity in embedding space rather than lexical overlap. Consequently, the metric measures distributional grounding in knowledge source. The fidelity is calculated in the following manner - 

Let the generated document be segmented into sentence-level units:

\[
S = \{s_1, s_2, \dots, s_T\}.
\]

For each segment $s_t$, we compute its embedding and compare it against the embeddings of all retrieved document chunks. The chunk with the highest cosine similarity is selected:

\[
c^*(s_t) =
\arg\max_{c \in \mathcal{C}}
\cos(\phi(s_t), \phi(c)),
\]

where $\phi(\cdot)$ denotes the embedding function and $\mathcal{C}$ is the set of retrieved chunks. 

If the highest similarity falls below a threshold $\tau$ (= 0.6~\cite{es-etal-2024-ragas}), the segment is labeled \textit{synthesized}, indicating that it cannot be confidently grounded in any retrieved source. This allows the metric to explicitly account for content generated through model synthesis rather than direct source support.

Each segment is then assigned to the KB of its matched chunk.

Let $w_i$ denote the total number of words in segments attributed to KB $i$, and let $w_{\text{synth}}$ denote the total number of words labeled synthesized. 

The realized distribution over KB usage is:

\[
Q_i =
\frac{w_i}
{\sum_{j=1}^{K} w_j + w_{\text{synth}}}
\quad
\text{for } i = 1,\dots,K
\]

\[
and Q_{\text{synth}} =
\frac{w_{\text{synth}}}
{\sum_{j=1}^{K} w_j + w_{\text{synth}}}.
\]

This yields a fine-grained estimate of how much of the generated document is actually supported by each KB.

(4) \textbf{\emph{Jensen--Shannon Divergence}} To quantify how closely realized KB usage matches prescribed weights, we compute the JS Divergence (JSD)~\cite{li2026attributingresponsecontextjensenshannon} between distributions $P'$ and $Q'$:

\[
P' = [P_1,\dots,P_K, 0]
\]

\[
Q' = [Q_1, \dots, Q_K, Q_{\text{synth}}],
\]

where the extra dimension accounts for synthesized content ($P_K=0$). JSD is defined as

\[
\text{JSD}(P' \parallel Q') = \frac{1}{2}\text{KL}(P' \parallel M) + \frac{1}{2}\text{KL}(Q' \parallel M), 
\]
\[
where \quad M = \frac{1}{2}(P' + Q'),
\]

$\text{KL}$ denoting Kullback--Leibler divergence. Bounded in $[0, \ln 2]$, we get a fidelity score in $[0,1]$:

\[
\text{Fidelity} = 1 - \frac{\text{JSD}(P' \parallel Q')}{\ln 2},
\]

where $1$ indicates perfect source fidelity and $0$ maximal divergence.

\paragraph{Large Language Models}
To ensure that observed differences are attributable to retrieval rather than generation, all retrieval-generation pipelines were run using the same backbone model, GPT-4.1~\cite{openai_gpt4_1}, accessed through the Azure AI Search~\cite{azure_ai_search_docs}. 

For automatic evaluation of Requirement Satisfaction and Context Relevancy, we use Claude Sonnet 4.6 ~\cite{anthropic_claude_sonnet_4_6} as the LLM-as-a-judge; these judgments are manually validated before analysis. All prompts for the LLMs, can be found in ~\ref{sec:judge-prompts}

\subsection{Results}

\paragraph{Generated Document Quality and Retrieval Effectiveness}

% Please add the following required packages to your document preamble:
% \usepackage{multirow}
% \usepackage{graphicx}
\begin{table*}[]
\resizebox{\textwidth}{!}{%
\begin{tabular}{|c|l|ll|ll|ll|ll|}
\hline
\textbf{} &
  \multicolumn{1}{c|}{\textbf{}} &
  \multicolumn{2}{c|}{\textbf{AI Technology \& Digital Media}} &
  \multicolumn{2}{c|}{\textbf{Climate \& Sustainability}} &
  \multicolumn{2}{c|}{\textbf{Healthcare \& Telehealth}} &
  \multicolumn{2}{c|}{\textbf{Finance}} \\ \hline
\textbf{Document Type} &
  \multicolumn{1}{c|}{\textbf{RAG Algorithm}} &
  \multicolumn{1}{c|}{\textbf{\begin{tabular}[c]{@{}c@{}}Requirement\\  Satisfaction (\%)\end{tabular}}} &
  \multicolumn{1}{c|}{\textbf{\begin{tabular}[c]{@{}c@{}}Context \\ Relevancy(\%)\end{tabular}}} &
  \multicolumn{1}{c|}{\textbf{\begin{tabular}[c]{@{}c@{}}Requirement\\  Satisfaction(\%)\end{tabular}}} &
  \multicolumn{1}{c|}{\textbf{\begin{tabular}[c]{@{}c@{}}Context \\ Relevancy(\%)\end{tabular}}} &
  \multicolumn{1}{c|}{\textbf{\begin{tabular}[c]{@{}c@{}}Requirement\\  Satisfaction(\%)\end{tabular}}} &
  \multicolumn{1}{c|}{\textbf{\begin{tabular}[c]{@{}c@{}}Context \\ Relevancy(\%)\end{tabular}}} &
  \multicolumn{1}{c|}{\textbf{\begin{tabular}[c]{@{}c@{}}Requirement\\  Satisfaction(\%)\end{tabular}}} &
  \multicolumn{1}{c|}{\textbf{\begin{tabular}[c]{@{}c@{}}Context \\ Relevancy(\%)\end{tabular}}} \\ \hline
\multirow{3}{*}{Policies} &
  Vanilla RAG &
  \multicolumn{1}{l|}{25} &
  57 &
  \multicolumn{1}{l|}{38} &
  32 &
  \multicolumn{1}{l|}{25} &
  30 &
  \multicolumn{1}{l|}{34} &
  41 \\ \cline{2-10} 
 &
  OG-RAG &
  \multicolumn{1}{l|}{34} &
  73 &
  \multicolumn{1}{l|}{68} &
  72 &
  \multicolumn{1}{l|}{35} &
  40 &
  \multicolumn{1}{l|}{51} &
  48 \\ \cline{2-10} 
 &
  \textbf{{\tool}} &
  \multicolumn{1}{l|}{\textbf{78}} &
  \textbf{76} &
  \multicolumn{1}{l|}{\textbf{76}} &
  \textbf{70} &
  \multicolumn{1}{l|}{\textbf{55}} &
  \textbf{70} &
  \multicolumn{1}{l|}{\textbf{60}} &
  \textbf{79} \\ \hline
\multirow{3}{*}{Product Launch} &
  Vanilla RAG &
  \multicolumn{1}{l|}{34} &
  45 &
  \multicolumn{1}{l|}{35} &
  40 &
  \multicolumn{1}{l|}{22} &
  28 &
  \multicolumn{1}{l|}{37} &
  40 \\ \cline{2-10} 
 &
  OG-RAG &
  \multicolumn{1}{l|}{38} &
  74 &
  \multicolumn{1}{l|}{56} &
  50 &
  \multicolumn{1}{l|}{33} &
  45 &
  \multicolumn{1}{l|}{5} &
  52 \\ \cline{2-10} 
 &
  \textbf{{\tool}} &
  \multicolumn{1}{l|}{\textbf{58}} &
  \textbf{78} &
  \multicolumn{1}{l|}{\textbf{74}} &
  \textbf{68} &
  \multicolumn{1}{l|}{\textbf{63}} &
  \textbf{71} &
  \multicolumn{1}{l|}{\textbf{61}} &
  \textbf{79} \\ \hline
\multirow{3}{*}{Merger \& Acquisitions} &
  Vanilla RAG &
  \multicolumn{1}{l|}{38} &
  52 &
  \multicolumn{1}{l|}{42} &
  38 &
  \multicolumn{1}{l|}{20} &
  32 &
  \multicolumn{1}{l|}{41} &
  45 \\ \cline{2-10} 
 &
  OG-RAG &
  \multicolumn{1}{l|}{39} &
  70 &
  \multicolumn{1}{l|}{54} &
  49 &
  \multicolumn{1}{l|}{32} &
  41 &
  \multicolumn{1}{l|}{5} &
  56 \\ \cline{2-10} 
 &
  \textbf{{\tool}} &
  \multicolumn{1}{l|}{\textbf{65}} &
  \textbf{78} &
  \multicolumn{1}{l|}{\textbf{83}} &
  \textbf{87} &
  \multicolumn{1}{l|}{\textbf{62}} &
  \textbf{68} &
  \multicolumn{1}{l|}{\textbf{73}} &
  \textbf{82} \\ \hline
\multirow{3}{*}{Academic Programs} &
  Vanilla RAG &
  \multicolumn{1}{l|}{28} &
  31 &
  \multicolumn{1}{l|}{40} &
  45 &
  \multicolumn{1}{l|}{27} &
  22 &
  \multicolumn{1}{l|}{34} &
  40 \\ \cline{2-10} 
 &
  OG-RAG &
  \multicolumn{1}{l|}{34} &
  64 &
  \multicolumn{1}{l|}{40} &
  44 &
  \multicolumn{1}{l|}{29} &
  42 &
  \multicolumn{1}{l|}{45} &
  52 \\ \cline{2-10} 
 &
  \textbf{{\tool}} &
  \multicolumn{1}{l|}{\textbf{72}} &
  \textbf{81} &
  \multicolumn{1}{l|}{\textbf{78}} &
  \textbf{82} &
  \multicolumn{1}{l|}{\textbf{67}} &
  \textbf{74} &
  \multicolumn{1}{l|}{\textbf{63}} &
  \textbf{7} \\ \hline
\end{tabular}%
}
\caption{Comparison of RAG algorithms across domains using Requirement Satisfaction and Context Relevancy}
\label{tab:average-across-domains}
\end{table*}

Before analyzing performance, we note that the LLM-as-a-judge used for scoring was manually validated and showed strong agreement with human verification, with an accuracy of 91\% for Requirement Satisfaction and 86\% for Context Relevancy. This suggests that the automatic scoring is reliable enough for comparative evaluation. Details on instructions for manual verification of scores is outlined in~\ref{sec:appendix-manual-validation}

Table 1 reports two complementary metrics: Requirement Satisfaction, which measures how much of the required content is covered in the generated enterprise document, and Context Relevancy, which measures how much of the retrieved evidence is relevant to the drafting task. Across all 16 settings, {\tool} yields substantially larger improvements: relative to OG-RAG, it improves Requirement Satisfaction by 58.1\% and Context Relevancy by 39.1\%; relative to Vanilla RAG, the gains are 109.2\% and 96.4\%, respectively. In practical terms, this means that {\tool} not only retrieves better supporting evidence, but also converts that evidence into drafts that satisfy far more of the required elements mentioned the input document.

A key pattern observed is that retrieval quality alone does not guarantee better document quality. For example, in AI Technology \& Digital Media product launch documents, OG-RAG improves Context Relevancy by 64.4\% over Vanilla RAG, but Requirement Satisfaction increases by only 11.8\%. W-RAG closes this gap, indicating that enterprise document generation depends not only on retrieving relevant passages, but also on composing evidence from multiple KBs in a source-aware manner. ~\ref{sec:appendix-analysis-rq1} details a fine grained analysis of the results.

\paragraph{Knowledge Base Influence \& Source Fidelity}

\begin{figure*}[t]
\centering

\begin{minipage}{0.248\textwidth}
\centering
\includegraphics[width=\linewidth]{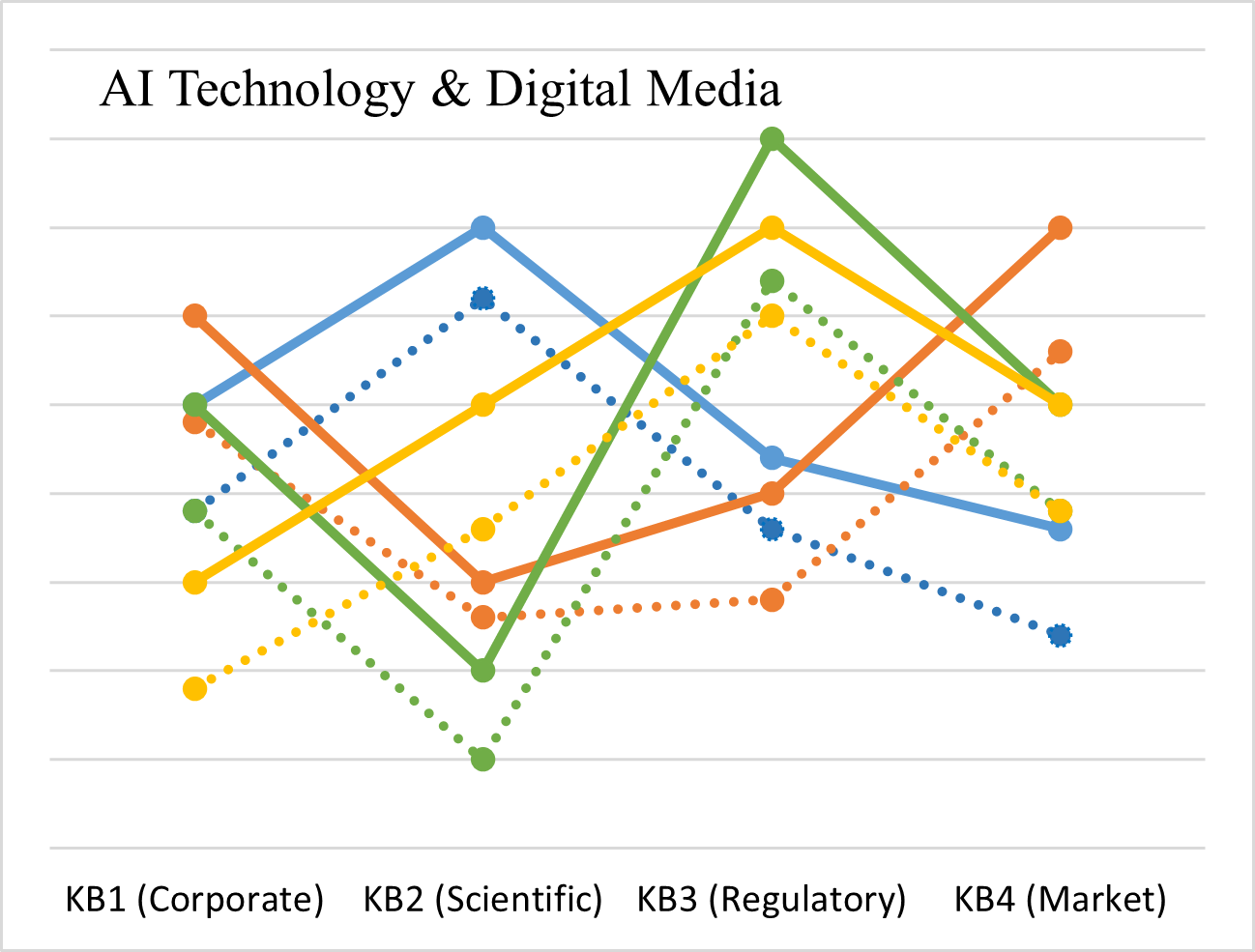}
% \subcaption{AI Technology}
\end{minipage}%
\begin{minipage}{0.248\textwidth}
\centering
\includegraphics[width=\linewidth]{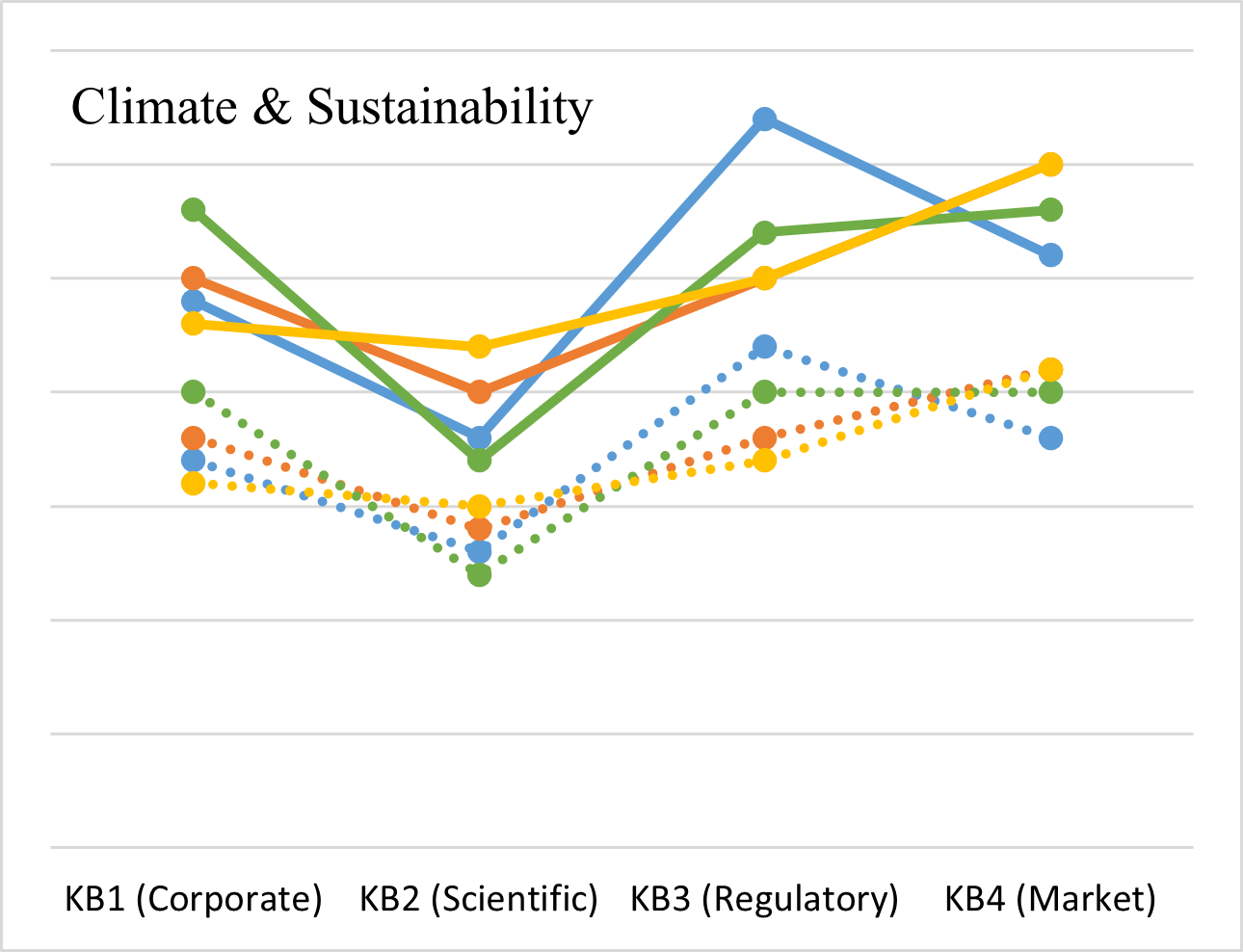}
% \subcaption{Climate}
\end{minipage}%
\begin{minipage}{0.248\textwidth}
\centering
\includegraphics[width=\linewidth]{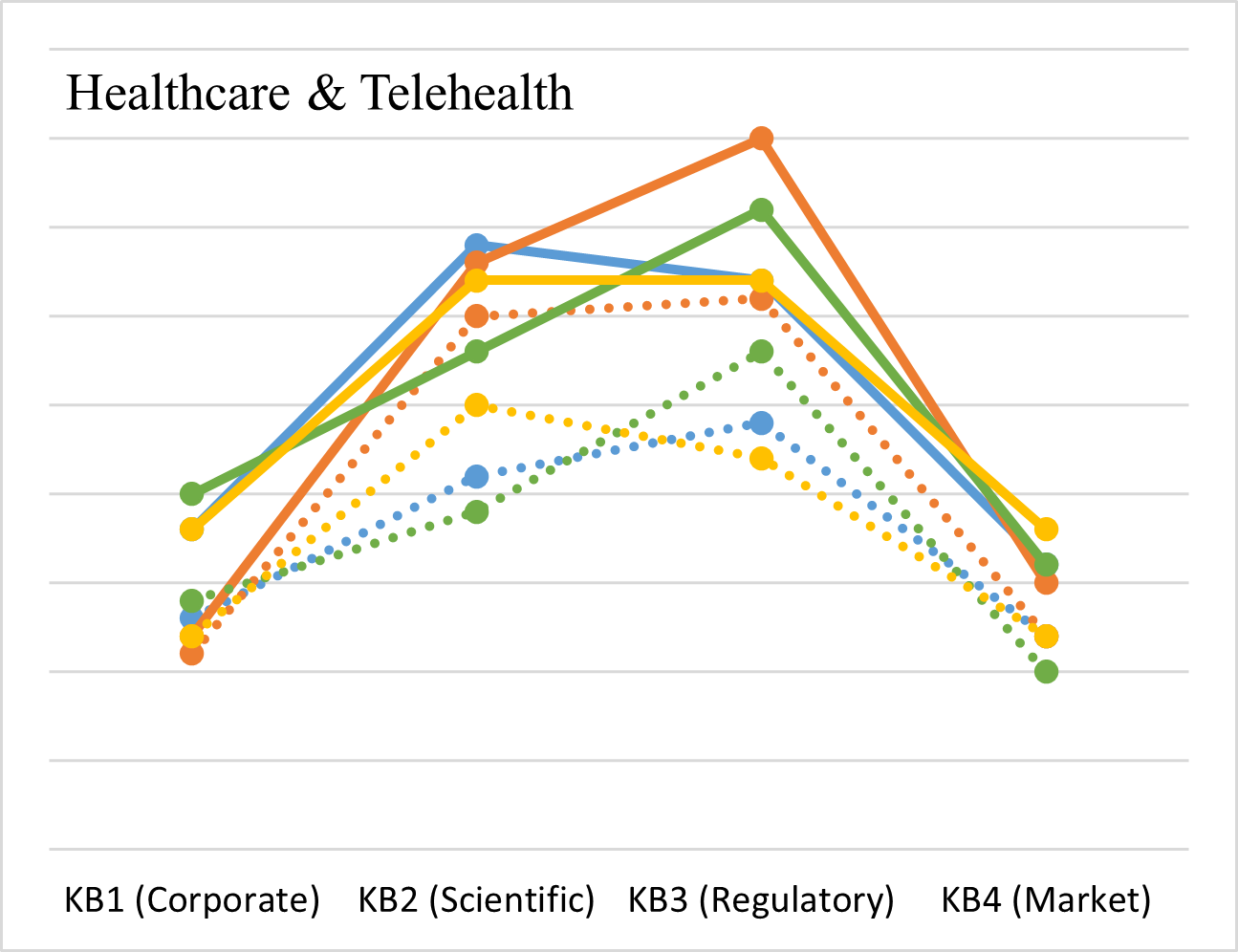}
% \subcaption{Healthcare}
\end{minipage}%
\begin{minipage}{0.248\textwidth}
\centering
\includegraphics[width=\linewidth]{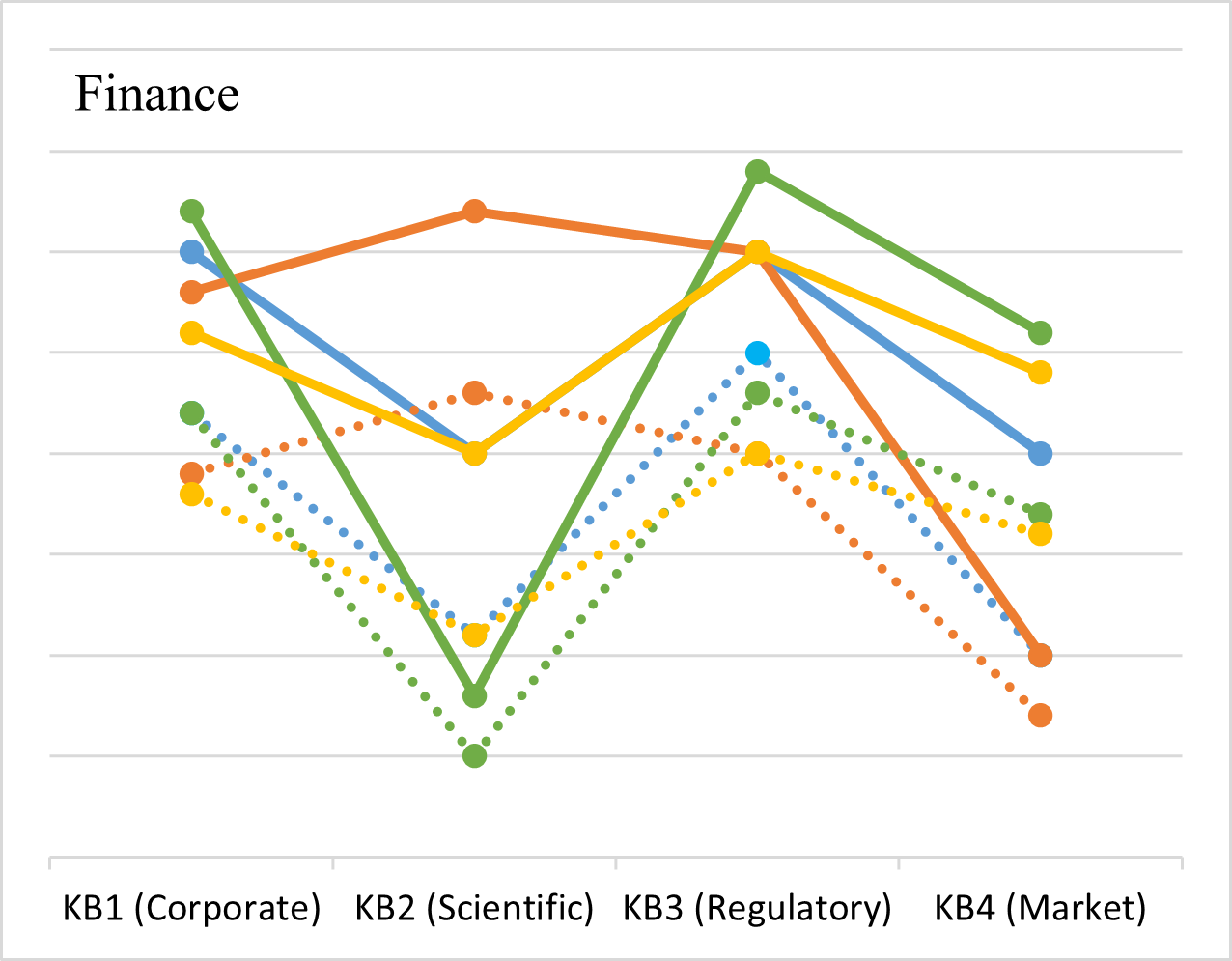}
% \subcaption{Finance}
\end{minipage}

\caption{Knowledge-base influence across domains. Solid lines denote prescribed KB weights at retrieval time, and dotted lines denote realized KB attribution in the generated output. Colors correspond to document types. Blue : Policies, Orange : Product Launch, Green : M\&A, Yellow: Academic Programs. Weights (Y-axis) range 0 to 50\%}
\label{fig:kb-distribution}
\end{figure*}

Figure~\ref{fig:kb-distribution} provides a domain-wise comparison between the prescribed source distribution at retrieval time and the realized KB attribution in the generated document. The degree to which the realized KB distribution tracks the input KB distribution serves as a direct visual indicator of source fidelity: when a KB's weight specified prior to retrieval is higher (i.e., more information is being retrieved from that KB than the others), the generated document reflects this intended source composition faithfully. The close alignment between the input and realized distributions across all domains suggests that the system maintains a strong linkage between retrieval-time source weighting and the resulting document content.

Consistent with the visual trends, the average JS divergence scores across domains range from 0.60 to 0.74, indicating moderate-to-strong agreement between prescribed and realized KB usage. This suggests that the generation process preserves the intended mixture of corporate, scientific, regulatory, and market knowledge when constructing enterprise documents. To move beyond visual evidence alone, Section~\ref{sec:kb-case-study} presents a controlled weight-override study in which prescribed KB mixtures are manually perturbed.

% Across domains, the realized distributions generally follow the same relative ordering as the prescribed inputs, indicating that the generation process meaningfully incorporates the weighted retrieval signals. 

\paragraph{Use Case: Corporate Proposal Drafting.}

A motivating enterprise use case for our setting is corporate proposal drafting, where proposal managers, solution architects, and business development teams prepare responses to complex RFPs for U.S. civilian and DoD agencies. These documents are highly structured, and heavily dependent on retrieving evidence from heterogeneous internal knowledge bases, including past proposals, contracts, capability decks, compliance documents, pricing artifacts, and agency-specific materials. In an internal retrieval-assisted drafting workflow used from 2025--2026, teams reported substantial productivity benefits, suggesting that proposal drafting is an important real-world testbed for multi-KB document generation. We provide a detailed case-study description in Appendix~\ref{sec:proposal-case-study}.

%% file: sections/conclusion.tex
\section{Conclusion}
\label{sec:conclusion}

This work introduced {\tool}, a retrieval framework designed for enterprise document generation tasks that require evidence from multiple heterogeneous knowledge bases. Our experiments show that global similarity ranking can lead to imbalanced context composition, where dominant sources overshadow others that are necessary for satisfying complex requirements. By combining ontology-guided topic extraction with source-aware context allocation, {\tool} improves both requirement satisfaction and context relevance on our benchmark dataset. These results suggest that, in multi-source enterprise settings, explicitly managing how retrieved context is distributed across knowledge bases can lead to more faithful and complete document generation.

% We presented \textsc{{\tool}}, a source-aware retrieval framework for enterprise document generation over heterogeneous knowledge bases. Unlike standard RAG pipelines that retrieve and rank evidence globally, \textsc{{\tool}} combines ontology-guided retrieval, local per-KB ranking, and weighted token allocation to better preserve the source composition required by the target document. To evaluate this setting, we introduced a new dataset spanning multiple enterprise document types and domains, designed to test not only retrieval relevance but also whether systems compose evidence from the appropriate mix of sources.

% Experiments show that existing RAG baselines often retrieve relevant passages yet still underperform on enterprise document drafting, particularly on Requirement Satisfaction. In contrast, \textsc{{\tool}} consistently improves both document quality and source fidelity, indicating that, in multi-source enterprise settings, explicitly managing how retrieved context is distributed across knowledge bases can lead to more faithful and complete document generation. 

%% file: sections/limitations.tex
\section{Limitations and Threats to Validity}
\label{sec:limitations}

\paragraph{Dataset and Task Scope.}
Our evaluation is conducted on a relatively small, synthetic enterprise-style dataset. While designed to reflect realistic multi-source document generation scenarios, it may not fully capture the diversity, scale, and noise present in real-world enterprise environments.

\paragraph{Evaluation Metrics.}
We evaluate generated outputs using task-specific metrics such as requirement satisfaction and context relevancy, which involve partial subjectivity in their definition or measurement. These metrics may not fully capture all aspects of document quality, such as factual correctness, coherence, or usability in downstream workflows.

\paragraph{Baseline Comparisons.}
Our comparisons are limited to standard retrieval-augmented generation baselines. We do not include heavily tuned retrieval strategies, which may reduce the observed performance gap.

\paragraph{Generality of the Approach.}
W-RAG is evaluated primarily in an enterprise document generation setting with structured knowledge sources. Its effectiveness in other domains—such as open-domain question answering or less structured retrieval environments—remains to be validated.

%% file: sections/appendix.tex
\label{sections:appendix}

\subsection{Dataset Details}
\label{sec:appendix-dataset}

\subsubsection{Input Requirement Documents}
Each sample in our dataset begins with an input requirement document, which serves as the drafting specification for the target enterprise document. Unlike short-form prompts or question-answer pairs, these inputs are structured requirement artifacts that describe the organizational setting, business context, constraints, intended audience, and mandatory sections that must appear in the final output. In practice, they are designed to resemble the kinds of requirement documentation used in real enterprise drafting workflows, where a writer or policy owner typically begins from a formal brief, scope document, governance note, or internal request rather than from a single natural-language query.

For each of the four document types—Policies, Product Launch documents, M\&A due diligence reports, and Academic Program proposals—we created requirement documents that specify both the situational context and the expected structural elements of the final draft. Document design was informed by common enterprise writing patterns, public institutional documentation, and the professional conventions associated with each document type. The goal was to create representative requirement specifications that capture the kinds of constraints, cross-functional evidence needs, and section-level expectations that arise in real-world enterprise authoring tasks.

Figure~\ref{fig:sample-input-policy} shows a sample input document for a corporate policy in the AI Technology Domain.

\begin{figure*}[t]
\centering
\fbox{
\parbox{0.96\textwidth}{
\small
\textbf{Enterprise Generative AI Governance Policy Requirement Documentation}

The report must be a structured document with the following four required sections:

\begin{enumerate}
    \item \textbf{Scope, Objectives \& Governance}
    \begin{itemize}
        \item Policy scope: covered systems (LLM-based assistants, prompt interfaces, orchestration layer), business units, and geographies
        \item Objectives: productivity enhancement, risk reduction versus uncontrolled usage, alignment with risk appetite
        \item Governance structure: AI Risk Committee, roles of Group Risk, IT, Data, Legal/Compliance, and Business Owners
        \item Policy ownership, review cycle (e.g., annual or on major technology/regulatory changes), and escalation paths
        \item Relationship to existing policies: Information Security, Model Risk, Outsourcing/Third Parties, Data Protection, Conduct \& Ethics
    \end{itemize}

    \item \textbf{Data, Privacy \& Security Controls}
    \begin{itemize}
        \item Data classification rules for prompts and outputs (e.g., prohibitions on secrets, PII, trading strategies, client identifiers in external LLMs)
        \item Allowed vs.\ disallowed data types by deployment type (on-prem vs.\ public cloud, single-tenant vs.\ shared)
        \item Data minimization and prompt hygiene standards; redaction, masking, and synthetic data where feasible
        \item Security controls: encryption, network segmentation, access control, logging, and monitoring
        \item Cross-border data transfer controls, including geofencing and regional hosting requirements
        \item Privacy and confidentiality requirements by jurisdiction (GDPR, UK GDPR, GLBA, local banking secrecy laws)
    \end{itemize}

    \item \textbf{Model Lifecycle, Quality \& Safety}
    \begin{itemize}
        \item Model sourcing and approval: criteria for selecting LLM providers; due diligence on training data, security, and reliability
        \item Model Risk Management alignment: classification of generative AI tools as models vs.\ tools; validation expectations for high-risk use cases
        \item Allowed vs.\ prohibited use categories (e.g., no unsupervised credit decisions, legal advice, investment recommendations, or regulatory reporting)
        \item Human-in-the-loop and human-on-the-loop requirements for business-critical use cases
        \item Hallucination / output reliability controls: mandatory citation of sources for internal knowledge tasks, disclaimers for outputs
        \item Testing, monitoring, and performance review: pre-deployment testing, ongoing monitoring, drift/failure modes, and incident thresholds
    \end{itemize}

    \item \textbf{Implementation, Monitoring \& Compliance}
    \begin{itemize}
        \item Onboarding process for new generative AI use cases: risk assessment templates, approvals, and documentation
        \item User access management: eligibility, training requirements, access revocation
        \item Training \& awareness: mandatory e-learning modules, acceptable use guidelines, role-specific training
        \item Monitoring \& reporting: centralized logging, dashboards for usage by business unit/region/use case, alerts for policy breaches
        \item Incident management: definition and handling of AI-related incidents (data leakage, harmful output, compliance breaches)
        \item Assurance \& audit: internal audit scope, regulatory exam readiness, evidence retention, KPI/KRI definitions
    \end{itemize}
\end{enumerate}
}
}
\caption{Sample input requirement document from the \textit{Policies} subset of the benchmark. The document specifies the required structure and mandatory content elements that the generated enterprise policy must cover.}
\label{fig:sample-input-policy}
\end{figure*}

\subsection{Ontology Construction, Topic Extraction, and Prompting}
\label{sec:appendix-ontology}

For the ontology-aware methods, we construct one ontology for each domain--document-type pair, yielding \textbf{16 ontologies} in total. Each ontology is built using a \textbf{semi-automated procedure}: an initial draft ontology is generated automatically from representative domain material and sample requirement documents, and is then manually reviewed, refined, and verified by domain experts. This process ensures that the final ontology captures the concepts, categories, and semantic relations most relevant to the corresponding enterprise document type.

An ontology in our setting is a formal representation of domain concepts and their relationships. Unlike a taxonomy or flat classification, it allows richer semantic organization by grouping related concepts under broader functional categories. For example, in a policy drafting ontology, concepts such as \textit{data privacy}, \textit{access control}, and \textit{audit logging} may be grouped under categories such as \textit{security}, \textit{governance}, and \textit{compliance}. 

The ontology is used in two places. First, it guides topic extraction from the input document, producing a structured topic representation of the drafting request. Second, it supports ontology-aware retrieval by enabling comparison between the input topics and the topics indexed from knowledge-base documents. To support this, each document added to a KB is processed through the same topic extraction pipeline, and the resulting topics are stored as document metadata.

Topic extraction is performed in two stages. In the first stage, the large language model identifies salient topic phrases from the document text. In the second stage, ontology content is supplied as additional context, so that extracted topics align more consistently with the ontology vocabulary and categories. The final output is mapped into the structured representation
\[
T = \{(t_i, g_i)\}_{i=1}^{N},
\]
where $t_i$ is an extracted topic phrase and $g_i$ is its ontology category.

When new documents are added to a KB, they do not affect retrieval immediately. They must first be re-indexed: topics are extracted from the new documents, the KB topic index is updated, and document counts and metadata are refreshed. Only after this process do newly added documents influence retrieval and source-weight estimation.

\subsection{LLM-as-a-Judge}
\label{sec:judge-prompts}

\subsubsection{Requirements Satisfaction Scoring}

For Requirement Satisfaction, the judge model is given three inputs: 

(1) the original requirement document,

(2) the generated enterprise document, and

(3) a document-type-specific requirement checklist prepared by a domain practitioner.

The judge is asked to evaluate the generated document item by item against the checklist and assign one of three labels for each requirement: fully satisfied, partially satisfied, or not satisfied. These are then converted into numeric scores according to the scoring rubric. The checklist is designed to capture the minimum essential elements that every valid document of that type should contain, independent of domain-specific style or wording. In the policy setting, example checklist items include whether the draft clearly defines scope, identifies a governance structure, specifies roles and responsibilities, states risk controls and escalation paths, defines monitoring and compliance mechanisms, and specifies a review cadence. A sample checklist is included in Figure~\ref{tab:policy-checklist} and a prompt example shown in Fig. \ref{fig:judge-prompt-rs}

\begin{table*}[t]
\centering
\small
\begin{tabular}{p{10cm}c}
\toprule
\textbf{Essential Item} & \textbf{Max Points} \\
\midrule
Scope clearly defined & 3 \\
Governance structure identified & 3 \\
Roles \& responsibilities assigned & 3 \\
Clinical/legal standards specified & 3 \\
Risk controls / escalation paths & 2 \\
Quality metrics defined & 2 \\
Workflows \& documentation rules & 2 \\
Approved technology specified & 2 \\
Training requirements & 2 \\
Monitoring \& compliance plan & 2 \\
Policy review cadence & 1 \\
\midrule
\textbf{Total} & \textbf{25} \\
\bottomrule
\end{tabular}
\caption{Sample requirement checklist used to score policy generation quality. Each item is scored as Full, Partial, or None, and then converted into a normalized Requirement Satisfaction score.}
\label{tab:policy-checklist}
\end{table*}

\begin{figure*}[t]
\centering
\fbox{
\parbox{0.96\textwidth}{
\small
\textbf{Sample LLM-as-a-Judge Prompt for Requirement Satisfaction}

You are evaluating whether a generated enterprise document satisfies the essential requirements of the requested document type.

You are given:

(1) The input requirement document

(2) The generated enterprise document

(3) A checklist of essential requirements

For each checklist item, assign one of the following labels:

(1) \textbf{Full}: the requirement is clearly and adequately satisfied

(2) \textbf{Partial}: the requirement is mentioned but incomplete or underspecified

(3) \textbf{None}: the requirement is absent or not meaningfully addressed

Convert these labels into scores:

(1) Full = full points

(2) Partial = half points

(3) None = zero points

Judge each checklist item independently. Focus on whether the required content is present, not on writing style or fluency. Return the item-level scores and the total score.
}
}
\caption{Prompt template used for LLM-based scoring of Requirement Satisfaction. Scores were manually verified by domain practitioners before analysis.}
\label{fig:judge-prompt-rs}
\end{figure*}

\subsubsection{Context Relevancy Scoring}

For Context Relevancy, the judge model is provided with:

(1) the input requirement document,

(2) the retrieved passages used as generation context, and

(3) instructions to label each retrieved passage as relevant or non-relevant to the drafting task.

The purpose of this prompt (shown in Fig.~\ref{fig:judge-prompt-cr}) is to isolate retrieval quality from final generation quality. A passage is judged relevant if it contributes useful evidence toward satisfying the requested enterprise document requirements, even if it is not copied directly into the final output. This is particularly important in our setting, where retrieved evidence may support governance, compliance, technical safeguards, or organizational procedures in complementary ways.

\begin{figure*}[t]
\centering
\fbox{
\parbox{0.96\textwidth}{
\small
\textbf{Sample LLM-as-a-Judge Prompt for Context Relevancy}

You are evaluating whether retrieved passages are relevant to an enterprise document drafting task.

You are given:
(1) The input requirement document
(2) A retrieved passage

Determine whether the passage is relevant to drafting the requested document.

Label the passage as:

(1) \textbf{Relevant}: the passage provides useful evidence for satisfying one or more required parts of the target document.

(2) \textbf{Not Relevant}: the passage is only loosely related, redundant, or does not materially help with the drafting task

Focus on task usefulness rather than general topical similarity. A passage may be relevant even if it supports only one section of the document.
}
}
\caption{Prompt template used for LLM-based scoring of Context Relevancy. Passage-level judgments were manually validated before analysis.}
\label{fig:judge-prompt-cr}
\end{figure*}

\subsection{Emperical Evaluation Results Analysis}

\begin{figure*}[t]
    \centering
    \includegraphics[width=\columnwidth]{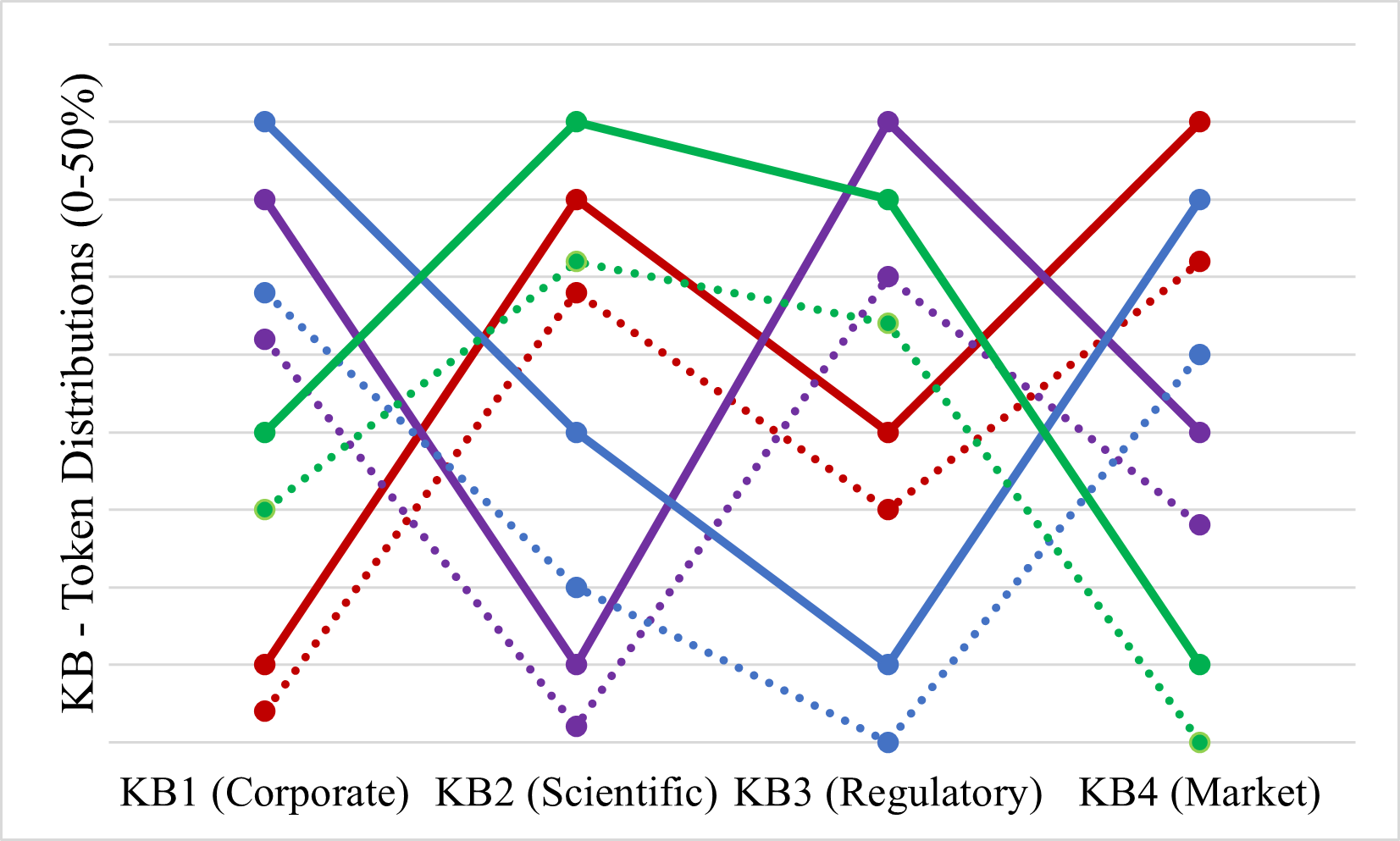}
    \caption{Controlled KB weight-override study on \textit{Policies} across all domains. Solid denotes one prescribed input KB distribution, constructed as a permutation of $(40,35,20,5)$, and the corresponding dotted curve denotes the realized KB attribution in the generated output.}
    \label{fig:kb-case-study}
\end{figure*}

\subsubsection{Fine-grained Analysis of Generated Document Quality and Retrieval Effectiveness}
\label{sec:appendix-analysis-rq1}

Table~\ref{tab:average-across-domains} reveals 2 robust patterns. First, \textsc{{\tool}} consistently outperforms both baselines on document quality, indicating that better retrieval alone is insufficient unless the retrieved evidence is also composed appropriately across KBs. Second, the magnitude of improvement is not uniform: document types that require broader synthesis across heterogeneous sources benefit the most from source-aware retrieval.

\paragraph{Policies.}
For \textit{Policies}, compared with \textsc{OG-RAG}, \textsc{{\tool}} yields a 43.1\% relative improvement in document quality and a 26.6\% improvement in retrieval quality. The strongest domain-level gain appears in \textit{AI Technology \& Digital Media}, where \textsc{Requirement Satisfaction} more than doubles relative to \textsc{OG-RAG} (34\%$\rightarrow$78\%). This suggests that policy drafting in technically evolving domains depends heavily on retrieving the right mixture of regulatory, organizational, and operational evidence. By contrast, \textit{Climate \& Sustainability} presents an informative counterexample: \textsc{OG-RAG} slightly outperforms \textsc{{\tool}} on \textsc{Context Relevancy} (72\% vs.\ 70\%), yet \textsc{{\tool}} still achieves higher \textsc{Requirement Satisfaction} (76\% vs.\ 68\%). This gap indicates that, for policy drafting, marginal improvements in passage-level relevance do not necessarily yield better documents unless the evidence is composed in a way that covers the required policy structure.

\paragraph{Product Launch.}
For \textit{Product Launch} documents, relative to \textsc{OG-RAG}, {\tool} observes an improvement of 44.6\% in document quality and 33.9\% in retrieval quality. The \textit{Healthcare \& Telehealth} setting is particularly revealing: \textsc{Requirement Satisfaction} rises from 33\% to 63\% and \textsc{Context Relevancy} from 45\% to 71\% when moving from \textsc{OG-RAG} to \textsc{{\tool}}. Product-launch drafting appears especially dependent on combining complementary evidence types---technical capabilities, market positioning, regulatory considerations, and organizational constraints---rather than merely retrieving thematically similar text. The relatively modest improvement of \textsc{OG-RAG} over vanilla RAG in \textsc{Requirement Satisfaction} suggest that identifying more relevant passages is not enough unless the system also retrieves them in a enterprise-appropriate mixture.

\paragraph{Merger \& Acquisitions.}
\textit{Merger \& Acquisitions} is the strongest-performing document type under \textsc{{\tool}} and exhibits some of the largest margins over both baselines. Averaged across domains, vanilla RAG reaches 35.3\% \textsc{Requirement Satisfaction} and 41.8\% \textsc{Context Relevancy}; \textsc{OG-RAG} improves to 43.8\% and 54.0\%; and \textsc{{\tool}} rises sharply to 70.8\% and 78.8\%. This translates into a 61.7\% relative gain in \textsc{Requirement Satisfaction} and a 45.8\% gain in \textsc{Context Relevancy} over \textsc{OG-RAG}. The largest single improvement occurs in \textit{Climate \& Sustainability}, where \textsc{{\tool}} reaches 83\% \textsc{Requirement Satisfaction} and 87\% \textsc{Context Relevancy}, compared with 54\% and 49\% under \textsc{OG-RAG}.

\paragraph{Academic Programs.}
\textit{Academic Programs}provides perhaps the clearest evidence that retrieval quality and document quality should be analyzed separately. Averaged across domains, vanilla RAG reaches 32.3\% \textsc{Requirement Satisfaction} and 34.5\% \textsc{Context Relevancy}; \textsc{OG-RAG} improves these to 37.0\% and 50.5\%; and \textsc{{\tool}} produces a much stronger 70.0\% and 76.8\%. Compared with \textsc{OG-RAG}, \textsc{{\tool}} improves \textsc{Requirement Satisfaction} by 89.2\% and \textsc{Context Relevancy} by 52.0\%.  The most striking example is \textit{Climate \& Sustainability}, where \textsc{OG-RAG} shows no improvement at all over vanilla RAG in \textsc{Requirement Satisfaction} (40\% vs.\ 40\%), yet \textsc{{\tool}} raises this score to 78\%. This pattern suggests that academic-program drafting is highly dependent on assembling evidence into a coherent, requirement-satisfying structure, and that source-balanced retrieval is particularly important when the target document spans policy, curriculum, compliance, and market-facing rationale.

\paragraph{Domain-level trends.}
Analysis across all document types reveals additional domain-specific structure. \textit{Healthcare \& Telehealth} is the most challenging domain overall: vanilla RAG averages only 23.5\% \textsc{Requirement Satisfaction} and 28.0\% \textsc{Context Relevancy}, while \textsc{OG-RAG} improves modestly to 32.3\% and 42.0\%. \textsc{{\tool}} substantially reduces this gap, improving the domain average to 61.8\% and 70.8\%. Relative to \textsc{OG-RAG}, this is a 91.5\% gain in document quality and a 68.5\% gain in retrieval quality, the largest relative improvement among all four domains. This suggests that healthcare drafting requires assembling fragmented and specialized evidence from multiple KBs and is therefore highly sensitive to source composition. In contrast, \textit{Climate \& Sustainability} attains the highest average \textsc{Requirement Satisfaction} under \textsc{{\tool}} at 77.8\%, followed by \textit{AI Technology \& Digital Media} at 68.3\% and \textit{Finance} at 64.3\%. Notably, \textit{AI Technology \& Digital Media} shows the highest retrieval scores already under \textsc{OG-RAG} (70.3\% average \textsc{Context Relevancy}), yet document quality remains much lower until weighting is introduced (36.3\% under \textsc{OG-RAG} vs.\ 68.3\% under \textsc{{\tool}}).

\paragraph{Retrieval quality versus document quality.}
A final observation is that \textsc{Context Relevancy} and \textsc{Requirement Satisfaction} are correlated but not interchangeable. OG-RAG frequently improves \textsc{Context Relevancy} by a large margin over vanilla RAG, yet document quality often improves only slightly. For example, in \textit{AI Technology \& Digital Media} \textit{Product Launch}, \textsc{Context Relevancy} increases from 29\% (45\% to 74\% ) when moving from vanilla RAG to OG-RAG, whereas \textsc{Requirement Satisfaction} rises only by 4\% (34\% to 38\%). Similar patterns appear in \textit{Policies} and \textit{Academic Programs}. This divergence supports the main claim of the paper: in enterprise multi-KB drafting, retrieving relevant passages is necessary but not sufficient. Systems must also regulate how much evidence from heterogeneous sources is retrieved and used during generation. \textsc{{\tool}} is effective precisely because it addresses both stages jointly.

\subsection{Fine-grained Analysis of Knowledge Base Influence}
\label{sec:appendix-analysis-rq1}
\paragraph{AI Technology \& Digital Media.}
This domain exhibits the largest mismatch between prescribed and realized KB usage, making it the most challenging setting for source control. In particular, realized attribution under weaker retrieval-generation settings appears to underuse KB3 (Regulatory) and KB4 (Market) relative to their prescribed contribution, while over- or under-shooting KB1 (Corporate) and KB2 (Scientific) depending on document type. This suggests that, in this domain, relevant evidence is more unevenly distributed across KBs, making generation more prone to source collapse. 

\paragraph{Climate \& Sustainability.}
Climate \& Sustainability shows the strongest agreement between prescribed and realized distributions. Across document types, the dotted curves largely track the shape of the solid curves, especially for KB3 (Regulatory) and KB4 (Market), which are both prominent contributors in this domain.

\paragraph{Healthcare \& Telehealth.}
In Healthcare \& Telehealth, the realized distributions capture the broad shape of the prescribed source mixture, but some attenuation is visible, particularly for KB3 (Regulatory) and KB4 (Market), where realized attribution is lower than intended for some document types. Even so, compared to weaker baselines, the realized source usage remains substantially better aligned with the prescribed mixture. 

\paragraph{Finance.}
Finance displays a mixed but still informative pattern. The realized distributions generally recover the main peaks in the prescribed source weights, especially for KB1 (Corporate) and KB3 (Regulatory), but show larger deviations for KB2 (Scientific) and KB4 (Market) in some document types. This suggests that finance-related generation is strongly influenced by corporate and regulatory evidence, while market-facing information is somewhat harder to preserve in the intended proportion. Nevertheless, the realized curves still follow the global structure of the prescribed distributions, indicating meaningful source influence rather than purely free-form generation.

% \paragraph{Overall interpretation.}
% Taken together, the figures show that the generated outputs are not source-agnostic: the contribution of each KB is reflected in the final text, and the degree of alignment can be measured through realized attribution. The main effect of \textsc{{\tool}} is therefore not merely to retrieve more relevant passages, but to ensure that the retrieved mixture of corporate, scientific, regulatory, and market evidence is more faithfully preserved during generation. This explains why improvements in source fidelity co-occur with higher document quality in the main results.

\subsubsection{KB Influence - Case Study}
\label{sec:kb-case-study}

To further test whether content generation is sensitive to the proportions of content retrieved from the KBs, we conduct a controlled weight-override study on the \textit{Policies} document type across all four domains. Rather than relying on the automatically recommended source weights, we manually specify four alternative input KB distributions for each example. Specifically, each document is run four times, using four permutations of the same weight set, ((40, 35, 20, 5)). In each run, a different KB is assigned the lowest weight ((5)), while the remaining three KBs receive the higher allocations ((40, 35,) and (20)). This design allows us to test whether systematically changing the prescribed source composition at retrieval time leads to corresponding changes in the realized KB usage in the generated document.

For each generated document, we compute the realized KB distribution using the sentence-level attribution procedure described in Section~\ref{sec:experiment-setup}, and then average the realized distributions across all policy examples. Figure\ref{fig:kb-case-study} shows the results. In each panel, the dots denote the actual prescribed and realized proportions for each KB, while the connecting lines are included only to make the distributional trend easier to compare.

The key result is that changing the prescribed KB mixture leads to corresponding changes in the realized source attribution of the generated document. In other words, when the retrieval context is reweighted to emphasize or de-emphasize a particular knowledge base, the content of the final enterprise document changes accordingly. This shows that enterprise document generation is highly sensitive to the type of information drawn from heterogeneous KBs: corporate, scientific, regulatory, and market sources do not contribute interchangeably, and shifting their relative presence in retrieval changes the informational composition of the output. 

\subsection{Corporate Proposal Drafting as a Target Application}
\label{sec:proposal-case-study}

We highlight corporate proposal drafting as a motivating enterprise application for weighted multi-KB retrieval-augmented generation. In many organizations, proposal teams prepare detailed responses to Requests for Proposals (RFPs) issued by government or commercial clients. These proposals are high-value documents: they must be persuasive, factually grounded, compliant with solicitation requirements, and tailored to the client, while also reflecting company capabilities, prior performance, pricing constraints, and organizational style. As a result, proposal drafting is both document-intensive and retrieval-dependent.

\paragraph{Why proposal drafting is challenging.}
Drafting a competitive proposal is time-consuming because relevant information is distributed across many heterogeneous internal sources. Writers must identify reusable content from prior proposal responses, past contracts, compliance artifacts, whitepapers, pricing documents, company capability decks, and agency-specific materials, then adapt that evidence into a coherent response. In practice, quality depends not only on retrieving relevant passages, but also on balancing evidence from multiple sources, maintaining factual consistency, satisfying compliance constraints, and customizing the draft for the target solicitation. These properties make proposal drafting a strong real-world instance of the broader problem studied in this paper.

\paragraph{Deployment context.}
The motivating workflow was observed in internal bids and proposals teams responsible for responses to U.S. civilian and Department of Defense solicitations. Users included approximately 15 internal personnel across three functional groups: proposal managers, technical or solution architects, and business development professionals. The teams worked on proposals spanning application development and modernization, IT support and service desk services, and cloud transformation and modernization. The internal knowledge sources used in this workflow included past proposal responses, prior contracts, capability decks, whitepapers and thought leadership documents, pricing materials, company compliance and regulatory documents, agency-specific materials, and graphics.

\paragraph{Workflow.}
The end-to-end drafting process is human-in-the-loop. A user begins by entering an RFP summary or proposal request. The retrieval system then identifies relevant internal materials from the available knowledge bases and drafts one or more proposal sections. The resulting draft is reviewed and edited by the proposal team, who refine language, add organization-specific positioning, adjust structure where necessary, and prepare the submission package. The system therefore functions as a drafting assistant rather than a fully autonomous proposal generator.

\paragraph{Observed operational utility.}
Between 2025 and 2026, the broader retrieval-assisted proposal drafting workflow was used in an ongoing internal setting. Users reported substantial efficiency benefits, including approximately 12--18 hours of time savings per team per week, with some reports indicating up to 20--25 hours saved during beta testing. Proposal turnaround was reported to be up to 3$\times$ faster, and first-draft preparation time was reduced by as much as 80\% in some cases. Draft acceptance was also high for early proposal drafts (``pink drafts''), with users reporting that 90--100\% of draft content was often acceptable as a starting point for downstream refinement. In addition, users reported relatively low editing burden for sections where the system could rely heavily on vetted internal knowledge and compliance-oriented document structure.

\paragraph{Most useful generated sections.}
Users found the system especially useful for sections that depend heavily on reusable institutional knowledge, including past performance, technical volumes, management volumes, and task-area content. They also reported that a complete Word document output was valuable operationally because it reduced manual setup effort associated with templates and structure selection. According to user feedback, the document structure was often highly compliant with solicitation requirements, which reduced formatting and compliance overhead during early-stage drafting.

\paragraph{Where human intervention remained necessary.}
Despite these benefits, expert review remained essential. Users consistently identified executive summaries, company differentiators, win themes, and staffing plans as sections requiring substantial human input. These components depend on competitive strategy, tacit organizational knowledge, dynamic personnel planning, and other forms of information that are often weakly represented in retrievable document collections. In particular, users noted that strong executive summaries and differentiator sections require persuasive positioning tailored to the solicitation, while staffing plans are too dynamic to be reliably delegated to an automated system.

\paragraph{Relation to \textsc{{\tool}}.}
It is important to distinguish the operational workflow above from the specific method proposed in this paper. The production-facing system described here was a retrieval-assisted drafting workflow, but \textsc{{\tool}} itself has not yet been fully deployed in production. Therefore, the operational efficiency numbers above should be interpreted as evidence that corporate proposal drafting is a high-value and practically relevant use case for retrieval-augmented generation, rather than as a direct causal evaluation of \textsc{{\tool}}. In this paper, we evaluate \textsc{{\tool}} offline on a proposal drafting dataset and show that weighted multi-KB retrieval improves generated document quality and evidence use relative to standard baselines. We view the real-world workflow described here as a motivating deployment context and our offline results as evidence that \textsc{{\tool}} is a promising next-step improvement for such settings.

\paragraph{Takeaway.}
Overall, the corporate proposal drafting workflow illustrates why weighted multi-KB retrieval matters in enterprise document generation. Proposal writing is not merely a search problem and not merely a text generation problem: it requires integrating heterogeneous internal evidence into a compliant, customizable, and organization-specific draft while preserving human control over final content. This makes it a natural target application for \textsc{{\tool}}.

\subsection{Manual Validation Instructions}
\label{sec:appendix-manual-validation}

To ensure evaluation reliability, all LLM-based Requirement Satisfaction and Context Relevancy scores were manually validated by 2 humans. Validators were instructed to review the generated document against the requirement checklist item by item, independently, using the following rubric: (1) \textbf{Yes} : the requirement is clearly and adequately addressed in the generated document; (2) \textbf{No} : the requirement is absent or insufficiently covered.

Validators were asked to focus on content coverage, not writing style. They were also instructed to judge each checklist item independently and not to reward broad but vague language unless the requirement was concretely addressed.